\documentclass[reqno]{amsart}

\usepackage{fullpage}

\usepackage{array,booktabs}

\usepackage{mathrsfs,textcomp,mathtools,accents}
\usepackage{hyperref}
\usepackage{bbm,eufrak}

\usepackage{enumerate}
\usepackage[shortlabels]{enumitem}
\setenumerate{label={\upshape(\roman*)},
  wide=0pt,topsep=0.2cm,labelsep=0em,itemsep=0.1cm,leftmargin=*}

\makeatletter
\@namedef{subjclassname@2020}{%
  \textup{2020} Mathematics Subject Classification}
\makeatother

\usepackage[dvipsnames]{xcolor}

\usepackage{tikz}
\usetikzlibrary{arrows,calc,positioning,spy,intersections}

\pgfdeclarelayer{background}
\pgfsetlayers{background,main}

\usepackage[font={scriptsize},labelfont={bf}, labelsep={period}]{caption}
\usepackage{mathdots}
\usepackage[figuresright]{rotating}
\usepackage{pdflscape}

\usepackage{fullpage}
\usepackage{graphicx}
\usepackage{fancyvrb}
\usepackage[latin1]{inputenc}
\usepackage[T1]{fontenc}
\usepackage[main=english]{babel}
\usepackage{wrapfig}
\usepackage{amsfonts}
\usepackage{amssymb}
\usepackage{amsmath}
\usepackage{eucal}
\usepackage{xcolor}
\usepackage{color}
\usepackage{float}
\usepackage{amsthm}
\usepackage{enumitem}
\usepackage{graphics} 
\usepackage{lmodern}
\usepackage[parfill]{parskip}
\usepackage{caption}
\usepackage{setspace}
\usepackage{listings}
\usepackage{verbatim}
\usepackage{hyperref}

\newtheorem{theorem}{Theorem}
\newtheorem{corollary}{Corollary}
\newtheorem{lemma}{Lemma}
\newtheorem{remark}[theorem]{Remark}

\newtheorem{definition}{Definition}
\newtheorem{proposition}[theorem]{Proposition}

\usepackage{pdflscape}

\begin{document}

\title{Asymptotic behaviour of the confidence region in orbit determination for hyperbolic maps with a parameter}%

\author{Nicola Bertozzi}%
\address{Dipartimento di Matematica, Universit\`a di Pisa, Largo Bruno
  Pontecorvo 5, 56127 Pisa, Italy} \email{nicola.bertozzi@phd.unipi.it}
  
  \author{Claudio Bonanno}%
\address{Dipartimento di Matematica, Universit\`a di Pisa, Largo Bruno
  Pontecorvo 5, 56127 Pisa, Italy} \email{claudio.bonanno@unipi.it}

\begin{abstract}
When dealing with an orbit determination problem, uncertainties naturally arise
from intrinsic errors related to observation devices and approximation models.
Following the least squares method and applying approximation schemes such
as the differential correction, uncertainties can be geometrically summarized in
confidence regions and estimated by confidence ellipsoids. We investigate the asymptotic behaviour of the confidence ellipsoids while the number of observations and the time span over which they are performed simultaneously increase. Numerical evidences suggest that, in the chaotic scenario, the uncertainties decay at different rates whether the orbit determination is set up to recover the initial conditions alone or along with a dynamical or kinematical parameter, while in the regular case there is no distinction. We show how to improve some of the results in \cite{maro.bonanno}, providing conditions that imply a non-faster-than-polynomial rate of decay in the chaotic case with the parameter, in accordance with the numerical experiments. We also apply these findings to well known examples of chaotic maps, such as piecewise expanding maps of the unit interval or affine hyperbolic toral transformations. We also discuss the applicability to intermittent maps.
\end{abstract}

\maketitle

\section{Introduction}\label{section:intro}

Orbit Determination problems have attracted a wide interest across the centuries, and still do. The core of modern Orbit Determination is the recovering of information about some unknown parameters related to a specific model starting from a set of previously acquired observations. Gauss' \textit{Least Squares Method} \cite{gauss} gave a remarkable contribution to Orbit Determination and is still largely employed within many modern algorithms related to impact monitoring activities or radio science experiments.\\
Various Orbit Determination problems are concerned with chaotic behaviour arising, for example, from close encounters of a celestial body with others being sufficiently massive. This emphasises the value of accurate predictions in impact monitoring activities \cite{milani.valsecchi} or space missions in which the spacecraft experiences close encounters with other celestial bodies in the Solar System \cite{lari.milani}.\\
The least squares method gives a description of the uncertainties through the so called \textit{confidence region} surrounding the nominal solution, which is the result of the application of the method which best approximates the real one. Hence, accuracy results could be inferred studying the evolution on the confidence region while the number of observations and, consequently, the timespan over which they are performed, increase.\\
Research in this direction has been enhanced by the numerical results in Serra et al. \cite{serra.spoto.milani}, Spoto and Milani \cite{spoto.milani} through the study of a model defined by the Chirikov standard map \cite{chirikov} depending on a parameter, which presents both ordered and chaotic regions. They simulated a set of observations by adding some noise to a real orbit of the map, then set up an orbit determination process in order to resume the true orbit. The numerical experiments highlighted a strong dependence of the decay of the uncertainties on the dynamics and on the nature of the parameters to be recovered. More specifically, if the true orbit was generated by an initial condition belonging to a chaotic zone, then the observed rate of decay of the uncertainties depends on whether the orbit determination is arranged to recover the initial conditions alone, in which case it is exponential, or together with an extra parameter, in which case the decay is polynomial. On the other hand, if the initial condition came from an ordered zone, then a polynomial rate could be observed in both the situations.\\
Analytical and formal proofs consistent with these numerical evidences were provided by S. Mar\`o and by S. Mar\`o and C. Bonanno in \cite{maro, maro.bonanno}, where hyperbolic transformations and a generalization of the standard map were taken as models for understanding, respectively, the chaotic and the ordered case.\\
In this paper we focus on the specific scenario of chaotic transformations depending on a parameter, and provide some conditions which, if satisfied by these maps, produce a polynomial-bounded decay of the uncertainties in an orbit determination process aiming to recover both the initial conditions and the varying parameter. Indeed, the results in \cite{maro.bonanno} concerning this case are in accordance with the numerical experiments, since a strictly slower than exponential rate of decay was shown, but an analytical proof of the precise polynomial decay is missing.\\
The rest of the paper is organized as follows. In Section \ref{section:settingandnotations} we present the setting and notations employed and give a formal description of the orbit determination problem and the least squares method, as outlined in \cite{milani.gronchi}. Section \ref{statement} specifies the problem we are interested in studying. In Section \ref{section:cd} we provide a condition that, if satisfied, imply the expected rate of decay, and list some examples. The same is done in Section \ref{section:hd}, where we point out a different condition which can be verified for other classes of maps. Sections \ref{section:proof_cd} and \ref{section:proofthrm2} are dedicated, respectively, to the proof of the main theorems \ref{thrm:cd} and \ref{thrm:hd}. In Section \ref{section:intermittent} we address the problems we may face dealing with intermittent maps, and give some results which still point towards the expected behaviour. In the concluding Section \ref{section:conclusions} we briefly summarize what we obtained and point out interesting directions that these studies may follow.

\section{Setting and Notations}\label{section:settingandnotations}

Let $X$ be a domain in $\mathbb{R}^d$ with $d \in \mathbb{N}^*$ and $K \subset \mathbb{R}$. Suppose that both $X$ and $K$ have non-empty interior. $X$ could be generalized to a differentiable Riemannian manifold employing slightly more intricate notations.\\
We consider a family of functions $\lbrace f_k : X \rightarrow X \rbrace_{k \in K}$ satisfying the following assumptions:
\begin{itemize}
\item For every $x \in X$, the map $k \mapsto f_k (x)$ from $K$ to $X$ is differentiable.
\item $X$ is endowed with a $\sigma$-algebra $\mathcal{B}$ such that $f_k$ is measurable and there exists an $f_k$-invariant probability measure $\mu_k$, absolutely continuous with respect to the Lebesgue measure, for every $k \in K$.
\item For every $k \in K$ and for $\mu_k$-a.e. $x \in X$, the function $f_k$ is differentiable with respect to $x$, with Jacobian matrix denoted by $F_k (x)$ and assumed to be invertible.
\item For every $k \in K$ and for $\mu_k$-a.e. $x \in X$, the function $f_k$ is also differentiable with respect to the couple $(x, k)$, with $\widetilde{F}_k(x)$ denoting the Jacobian matrix.
\item $\exists \, M \in \mathbb{R}^+$ s. t. $\sup\limits_{x \in X} \left| \left| \dfrac{\partial f_k}{\partial k}(x) \right| \right| \le M$, for a given.
\end{itemize}
\phantom{bla}\\
Given $n \in \mathbb{N}$, we call $f_k^n$ the $n-$th iterate of the map, where $f_k^0:= id_X$ is the identity. Its Jacobian matrix with respect to $x \in X$ can be expressed by the chain rule as
\begin{center}
$F_k^n(x) = \begin{cases}
F_k(f_k^{n-1}(x)) F_k(f_k^{n-2}(x)) \dots F_k (x) & \text{for \ } n \ge 1,\\
I_{d \times d} & \text{for \ } n=0.
\end{cases}
$
\end{center}
In a similar way, the Jacobian matrix of $f_k^n$ with respect to $(x, k)$ will be called
\begin{center}
$\widetilde{F}_k^n (x) := \begin{bmatrix}
F_k^n (x) & \bigg| & \dfrac{\partial f_k^n}{\partial k}(x)
\end{bmatrix}$.
\end{center}

In order to simplify the discussion, we introduce the following definition:
\begin{definition}
\ \\
For a fixed $k \in K$, we say that a point $x \in X$ is \textbf{suitable} if, for every $n \in \mathbb{N}$, $F_k^n (x)$ is well defined and invertible.\\
A pair $(x , k) \in X \times K$ is called \textbf{suitable} if $x$ is suitable for the map $f_k$.
\end{definition}
Adopting the approach of Milani and Gronchi \cite{milani.gronchi}, we set up an orbit determination process modelled by $f_k$: we assume to have a set $\lbrace \overline{X}_n \rbrace_{n=0}^N$ of observations performed at times $0,1,\dots , N$ and related to a system whose evolution is represented by the orbits of $f_k$. Our aim is to recover some unknown parameters which characterize this map, namely the initial conditions and the value of $k$. The least squares method (if convergent) will provide an initial state $x^* \in X$ and a specific $k^* \in K$ so that the first $N$ terms of the orbit of $x^*$ under $f_{k^*}$ will be a good approximation of the observations.\\
Firstly, we define the \textit{residuals}
\begin{align*}
\widetilde{\xi}_n (x,k) := \overline{X}_n - f_k^n(x),
\end{align*}
where $(x,k) \in X \times K$ is a suitable pair and $n \in \lbrace 0, 1, \dots , N \rbrace$.\\
The \textit{nominal solution} $(x^*,k^*)$ is chosen as the minimum point of the \textit{target function}
\begin{align*}
\widetilde{Q}(x,k) := \dfrac{1}{N+1} \sum\limits_{n=0}^N \widetilde{\xi}_n (x,k)^T \widetilde{\xi}_n (x,k) = \dfrac{1}{N+1} \sum\limits_{n=0}^N \parallel \widetilde{\xi}_n (x,k) \parallel_2^2.
\end{align*}
Finding the minima of this function (if they even exist) is all but a trivial task, usually processed using iterative schemes such as the Gauss-Newton algorithm and the differential corrections \cite{milani.gronchi}. Proving the existence and computing the nominal solutions will not be of our concern since we will focus on maps that can be dealt with standard or advanced techniques such as the multi-arc approach \cite{serra.spoto.milani}.\\
\\
In general, the nominal solution $x^*$ differs from the real one, due to the intrinsic errors coming from the observation process. Therefore, one considers acceptable the elements of the \textit{confidence region}
\begin{center}
$\widetilde{Z}(\sigma): = \left\{ (x,k) \in X \times K \, : \, \widetilde{Q}(x,k) \le \widetilde{Q}(x^*,k^*) + \dfrac{\sigma^2}{N+1} \right\}$,
\end{center}
the set of couples $(x,k)$ on which the target function takes values which are slightly bigger than the minimum depending on an empirical parameter $\sigma$. The parameter $\sigma$ depends on the statistical properties of the specific problem  and without loss of generality we can normalize it to $\sigma = 1$.\\
If we expand the target function in a neighbourhood of the nominal solution up to the second order we find that
\begin{align*}
\widetilde{Q}(x,k)\simeq \widetilde{Q}(x^*,k^*)+\dfrac{\partial \widetilde{Q}}{\partial (x,k)}(x^*,k^*)\begin{bmatrix}
x-x^* \\ k-k^*
\end{bmatrix} + \dfrac{1}{2} \begin{bmatrix}
x-x^* \\ k-k^*
\end{bmatrix}^T \cfrac{\partial^2 \widetilde{Q}}{\partial (x,k)^2}(x^*,k^*) \begin{bmatrix}
x-x^* \\ k-k^*
\end{bmatrix}.
\end{align*}

Being $(x^*,k^*)$ a minimum point of $\widetilde{Q}$, and assuming that the residuals are small (that is, the least squares method worked out well), we can neglect higher order terms and approximate the target function as
\begin{align*}
\widetilde{Q}(x,k) \simeq \widetilde{Q}(x^*,k^*) + \dfrac{1}{N+1} \begin{bmatrix}
x-x^* \\ k-k^*
\end{bmatrix}^T \left[ \sum\limits_{n=0}^N \widetilde{F}_{k^*}(x^*)^T \widetilde{F}_{k^*}(x^*) \right] \begin{bmatrix}
x-x^* \\ k-k^*
\end{bmatrix}.
\end{align*}
We define the \textit{normal matrix} as
\begin{align*}
\widetilde{C}_N(x^*,k^*) := \sum\limits_{n=0}^N \widetilde{F}_{k^*}(x^*)^T \widetilde{F}_{k^*}(x^*),
\end{align*}
and the \textit{covariance matrix} as its inverse
\begin{align*}
\widetilde{\Gamma}_N(x^*,k^*) := \left[ \widetilde{C}_N(x^*,k^*) \right]^{-1}.
\end{align*}
We observe that these matrices are symmetric and that, since in our set up the Jacobian matrix $\widetilde{F}_k(x)$ has full rank, they are positive definite.\\
Therefore, we can now employ the \textit{confidence ellipsoid}
\begin{center}
$\widetilde{\mathcal{E}}_N (x^*, k^* ):= \Bigg\{ (x,k) \in X \times K \, : \, (x,k) \text{ is suitable and}\begin{bmatrix} x-x^* \\ k-k^* \end{bmatrix}^T \widetilde{C}_N(x,k) \begin{bmatrix}
x-x^* \\ k-k^*
\end{bmatrix} \le 1 \Bigg\}$
\end{center}
as a relatively close description of the confidence region.\\
The size of the ellipsoid is determined by the covariance matrix: calling 
\begin{align*}
\widetilde{\lambda}_N^{(1)}(x,k) \le \widetilde{\lambda}_N^{(2)}(x,k)\le \dots \le \widetilde{\lambda}_N^{(d+1)}(x,k)
\end{align*}
the eigenvalues of $\widetilde{\Gamma}_N (x,k)$ (all positive because the matrix is positive definite), a standard geometric fact tells us that the axes of $\widetilde{\mathcal{E}}_N(x^*,k^*)$ have length
\begin{center}
$2 \sqrt{\widetilde{\lambda}_N^{(j)} (x^*,k^*)}, \text{where} \  j \in \lbrace 1,2, \dots , d+1 \rbrace $.
\end{center}

\section{Statement of the Problem}\label{statement}

Accuracy is an essential issue in orbit determination. Since the confidence ellipsoid approximates the confidence region, we can use it to outline the inevitable errors arising from the observation process. In particular, it is interesting to examine its behaviour while the number of observations and, consequently, the timespan over which they are performed increase. In general, it is reasonable to expect that, gathering more information, the orbit determination will be more precise, hence the axes of the ellipsoid will shrink.\\
The idea of studying the eigenvalues of the covariance matrix (strictly related to the size of the confidence ellipsoid, as seen in the previous section) was conducted by Milani et al. \cite{serra.spoto.milani, spoto.milani}, who worked on a model problem based on the Chirikov standard map. Their numerical results show that, if the orbit determination is set up in an ordered environment, the uncertainties decay at a polynomial rate, while in the chaotic scenario the rate is exponential if the unknown parameters to be recovered are the initial conditions alone, but polynomial when another parameter ($k$ in the previous section) is added to the unknown to be determined.\\
These facts were formally proved by Mar\`o and Bonanno \cite{maro.bonanno} for chaotic maps and by Mar\`o \cite{maro} for a generalization of the standard map. However, in the case of chaotic maps with unknown parameters including $k$, the results show that the rate of decay of the uncertainties is strictly slower than any exponential, but there are no hints about a specific polynomial estimate. Hence, in this paper we study the following\\
\\
\textbf{Problem}\\
To estimate the rate of decay of the uncertainties with a polynomial bound in the case of chaotic maps with unknown parameters including both initial conditions and $k$.\\
\\
We look for appropriate conditions which imply a polynomial bound from below on the maximum eigenvalue of the covariance matrix, so that we can infer that the greatest axis of the confidence ellipsoid decays with a rate which is not faster than a polynomial.\\
For every suitable pair $(x,k) \in X \times K$, let us (formally) call
\begin{align*}
S_k(x):= \sum\limits_{i=1}^{+\infty} \parallel [F_k^i(x)]^{-1} \parallel.
\end{align*}
In the next sections we introduce conditions on $S_k(x)$ that, if satisfied by the map $f_k$, imply the requested polynomial bound on the greatest eigenvalue $\widetilde{\lambda}_N^{(d+1)}(x,k)$ of the covariance matrix $\widetilde{\Gamma}_N(x,k)$. Then, we provide some examples of well-known hyperbolic maps satisfying the conditions.\\
The proofs of the main theorems are in \autoref{section:proof_cd} and \autoref{section:proofthrm2}.

\section{Condition $C_d$ and Applications}\label{section:cd}

We say that \textit{condition $C_d$} holds if $\forall \, k \in K \ \exists \, \sigma_k \in \mathbb{R}^+$ s.t. if $(x,k)\in X \times K$ is suitable then $S_k(x) \le \sigma_k$.\\
Here the subscript $d$ in the notation $C_d$ refers to the dimension of the domain $X$.

\begin{theorem}
\label{thrm:cd}
\ \\
Let $f_k : X \rightarrow X$ be such that condition $C_d$ is verified. Then, for a fixed suitable pair $(x,k) \in X \times K$ and for every $N \in \mathbb{N}^*$, the greatest eigenvalue $\widetilde{\lambda}_N^{(d+1)}(x,k)$ of the covariance matrix $\widetilde{\Gamma}_N (x,k)$ satisfies
\begin{align*}
\widetilde{\lambda}_N^{(d+1)} (x,k) \ge \dfrac{1}{\left[M \sigma_k\right]^2} \cdot \dfrac{1}{(N+1)}.
\end{align*}
\end{theorem}

A straightforward application of this result concerns the broadly studied uniform piecewise expanding maps of the unit interval. More specifically, let us consider a family $\lbrace f_k: [0,1] \rightarrow [0,1] \rbrace_{k \in K}$ such that, for every $k \in K$, we have that:

\begin{itemize}
\item There exists a sequence $0=s_0 < s_1 < \dots < s_{m_k} =1$, with $m_k \in \mathbb{N}^* \cup \lbrace +\infty \rbrace $, such that, naming $I_j := (s_{j-1} , s_j )$ for every $j = 1, 2, \dots , m_k$, we have that $f_k \vert_{I_j}$ is of class $C^2$ and has a $C^1$ extension to the closure $\overline{I}_j$.
\item There exists $c_k \in (1, + \infty)$ such that, for all $j = 1, \dots , m_k$, we have $\inf\limits_{x \in I_j} |(f_k \vert_{I_j})' (x) | \ge c_k$.
\end{itemize}

\begin{corollary}
\ \\
A family $\lbrace f_k : [0,1] \rightarrow [0,1] \rbrace_{k \in K}$ of piecewise expanding maps of $[0,1]$ satisfies condition $C_1$.
\end{corollary}

\begin{proof}
By assumption, we know that, for every suitable pair $(x,k) \in [0,1] \times K$,
\begin{align*}
|f_k'(x)| \ge c_k > 1.
\end{align*}
Hence, whenever $i \in \mathbb{N}^*$, we have that
\begin{align*}
|(f_k^i)'(x)|= \prod\limits_{j=0}^{i-1} |f_k'(f_k^{j}(x))| \ge \prod\limits_{j=0}^{i-1} c_k = c_k^i,
\end{align*}
and thus
\begin{align*}
\sum\limits_{i=1}^{+\infty} | (f_k^i)'(x)^{-1}| = \sum\limits_{i=1}^{+\infty}  |(f_k^i)'(x)|^{-1} \le  \sum\limits_{i=1}^{+\infty} c_k^{-i} = \dfrac{1}{c_k - 1}. 
\end{align*}
Condition $C_1$ holds choosing $\sigma_k = \dfrac{1}{c_k - 1}.$
\end{proof}
%\ \\
%\vspace{-0.3cm}

\section{Condition $H_d$ and Applications}\label{section:hd}
%\ \\

\textit{Condition $H_d$} is said to be verified if, for fixed $k \in K$ and $p,q \in \mathbb{N}$ such that $p+q=d$, the following is true for every suitable $x \in X$:
\begin{itemize}
\item$\exists V_k \in \mathbb{R}^{d\times d}$ such that
\begin{center}
$\begin{aligned}[t]
V_k^{-1} F_k(x) V_k = \left[ \begin{array}{c | c}
A_k(x) & 0_{p,q}\\[1.4ex]
\hline\\[-1.3ex]
0_{q,p} & B_k(x)
\end{array} \right] =: D_k(x)
\end{aligned}$
\end{center}
is a simultaneous block diagonalization for $F_k(x)$, where 
\begin{align*}
\begin{array}{l l}
A_k(x) \in \mathbb{R}^{p \times p}, \ \ & \ \ B_k(x) \in \mathbb{R}^{q \times q}.
\end{array}
\end{align*}
For every $n \in \mathbb{N}^*$, we will call
\begin{align*}
A_k^n(x):= \prod\limits_{j=0}^{n-1} A_k (f_k^j(x)) \ \ \ \ \text{and} \ \ \ \ A_k^0(x):= I_{p \times p};\\
B_k^n(x):= \prod\limits_{j=0}^{n-1} B_k (f_k^j(x)) \ \ \ \ \text{and} \ \ \ \ B_k^0(x):= I_{q \times q}.
\end{align*}
\item There exists $\alpha_k \in \mathbb{R}^+$ such that $\sum\limits_{i=1}^{+\infty} \parallel [A_k^i(x)]^{-1} \parallel_2 \le \alpha_k$.
\item There exists $\beta_k \in \mathbb{R}^+$ such that, for every $n \in \mathbb{N}^*$, $\sum\limits_{i=0}^{n-1} \parallel B_k^i(f_k^{n-i}(x)) \parallel_2 \le \beta_k$.
\end{itemize}

\begin{theorem}
\label{thrm:hd}
\ \\
Let $f_k: X \rightarrow X$ be such that condition $H_d$ is verified. Then, for every suitable pair $(x, k) \in X \times K$ and for every $N \in \mathbb{N}^*$, the greatest eigenvalue $\widetilde{\lambda}_N^{(d+1)}(x,k)$ of the covariance matrix $\widetilde{\Gamma}_N(x,k)$ satisfies
\begin{align*}
\widetilde{\lambda}_N^{(d+1)}(x,k) \, \ge \, \dfrac{\parallel \! V_k^{(p)} \!\cdot \! L(x,k) \! \parallel_2^2 + 1}{\left( M \! \parallel \! V_k \! \parallel_2 \parallel \! V_k^{-1} \! \parallel_2 \right)^2 (\alpha_k^2 + \beta_k^2)}\cdot \dfrac{1}{N+1}.
\end{align*}
\end{theorem}

This fact can be employed for showing a slight generalization of some results in \cite{maro.bonanno} about affine hyperbolic diffeomorphisms of the torus $\mathbb{T}^d$. In particular, we study maps of the form
\begin{equation} \label{eq:ahd}
\begin{array}{r c l c}
f_k: & \mathbb{T}^d & \rightarrow & \mathbb{T}^d\\
\ & x & \mapsto & P_k x + q_k,
\end{array}
\end{equation}
with $P_k \in SL(d, \mathbb{Z})$ without eigenvalues of modulus $1$, and $q_k \in \mathbb{R}^d$.\\
In order to show that the above maps satisfy condition $H_d$, we need a couple of lemmas.
\begin{definition}
\ \\
The \textbf{spectral radius} of a matrix $P \in \mathbb{C}^{d \times d}$ is defined as
\begin{align*}
\rho (P) := \max \lbrace | \nu | \, : \, \nu \ \text{is an eigevalue of } P \rbrace.
\end{align*}
\end{definition}

The following is a well known fact of linear algebra \cite[p. 119]{bini.capovani.menchi}.
\begin{lemma}
\ \\
For every matrix $P \in \mathbb{C}^{d \times d}$ and for every $\epsilon \in \mathbb{R}^+$ there exists a matrix norm $| \cdot |_\epsilon$ such that
\begin{align*}
| P |_\epsilon \le \rho (P) + \epsilon.
\end{align*}
\end{lemma}

\begin{lemma}
\label{lemma:contrazione}
\ \\
If $P \in \mathbb{C}^{d \times d}$ has spectral radius $\rho (P) < 1$, then there exist two constants, $c \in \mathbb{R}^+$ and $\theta \in (0,1)$, such that, for every $n \in \mathbb{N}$,
\begin{align*}
\parallel P^n \parallel_2 \, \le \, c \, \theta^n.
\end{align*}
\end{lemma}

\begin{proof}
By the hypothesis, we can fix an $\epsilon \in \mathbb{R}^+$ such that $\theta:= \rho ( P ) + \epsilon < 1$.

The previous lemma provides a matrix norm $| \cdot |_\epsilon$ with $| P |_\epsilon \le \theta.$\\
Moreover, all the matrix norms are equivalent, hence there exists a constant $c \in \mathbb{R}^+$ such that, for every matrix $Q \in \mathbb{C}^{d \times d}$, $\parallel Q \parallel_2 \le c \, | P |_\epsilon.$\\
Thus, we readily conclude that, for every $n \in \mathbb{N}$, $\parallel \! P^n \!\! \parallel_2 \le c \, | P^n |_\epsilon \le c \, | P |_\epsilon^n \le c \, \theta^n.$
\end{proof}

\begin{corollary}
\ \\
A family $\lbrace f_k : \mathbb{T}^d \rightarrow \mathbb{T}^d \rbrace_{k \in K}$ of affine hyperbolic toral diffeomorphisms satisfies condition $H_d$.
\end{corollary}

\begin{proof}
Let us consider such transformations with the same notations as in eq. \ref{eq:ahd}.\\
We know that, for every suitable $(x,k) \in \mathbb{T}^d \times K$ and for every $n \in \mathbb{N}$, the Jacobian matrices are constant: $F_k^n (x) = P_k^n$.\\
By assumption, we know that every eigenvalue of $P_k$ is non-vanishing and with modulus strictly smaller or strictly larger than $1$.\\
We call $E_k^u$ the sum of all the generalized eigenspaces for eigenvalues of modulus greater than $1$ and, similarly, $E_k^s$ the sum of all the generalized eigenspaces for eigenvalues of modulus less than $1$.\\
\\
If $dim(E_k^u) = p$ and $dim(E_k^s) = q$ (so that $p+q=d$), let $\lbrace \mathbf{v}_k^{(1)}, \dots ,\mathbf{v}_k^{(p)} \rbrace \subset \mathbb{R}^d$ be a basis for $E_k^u$, and $\lbrace \mathbf{v}_k^{(p+1)}, \dots ,\mathbf{v}_k^{(d)} \rbrace \subset \mathbb{R}^d$ be a basis for $E_k^s$.\\
We may assemble these vectors into two matrices:
\begin{align*}
V_k^u := \left[ \begin{array}{c | c | c}
\mathbf{v}_k^{(1)} & \dots & \mathbf{v}_k^{(p)}
\end{array} \right] \in \mathbb{R}^{d \times p} \ \ \ \ 
\text{and}\ \ \ \ 
V_k^s := \left[ \begin{array}{c | c | c}
\mathbf{v}_k^{(p+1)} & \dots & \mathbf{v}_k^{(d)}
\end{array}\right] \in \mathbb{R}^{d \times q}.
\end{align*}
\phantom{bla}\\
Now, being $E_k^u$ and $E_k^s$ invariant under the action of $P_k$, the matrix
\begin{align*}
V_k := \left[ \begin{array}{c | c}
V_k^u & V_k^s
\end{array} \right] \in \mathbb{R}^{d \times d}
\end{align*}
block-diagonalizes $P_k$ in the following way:
\begin{align*}
V_k^{-1} P_k V_k = \left[ \begin{array}{c | c}
A_k & 0_{p,q}\\[1.4ex]
\hline\\[-1.3ex]
0_{q,p} & B_k
\end{array} \right],
\end{align*}
where $A_k  \in \mathbb{R}^{p \times p}$ and $B_k \in \mathbb{R}^{q \times q}$.\\
\\
In order to verify condition $H_d$, we show that
\begin{itemize}
\item there exists $\alpha_k \in \mathbb{R}^+$ such that $\sum\limits_{i=1}^{+\infty} \parallel [A_k^i ]^{-1} \parallel_2 = \sum\limits_{i=1}^{+\infty} \parallel A_k^{-i} \parallel_2 \le \alpha_k$;\\
\item there exists $\beta_k \in \mathbb{R}^+$ such that $\sum\limits_{i=0}^{+\infty} \parallel B_k^i\parallel_2 \le \beta_k$ , for every $n \in \mathbb{N}^*$.
\end{itemize}

Now, by construction, all the eigenvalues of the matrix $A_k$ have modulus greater than $1$. Therefore, its inverse $A_k^{-1}$ is such that $\rho (A_k^{-1} ) < 1$,
because its eigenvalues are the reciprocals of those of $A_k$.\\
Lemma \ref{lemma:contrazione} implies that there exist two constants, $c_{k,p} \in \mathbb{R}^+$ and $\theta_{k,p} \in (0,1)$, such that, for every $i \in \mathbb{N}$, we have $\parallel P_k^{-i} \parallel_2 \le c_{k,p} \theta_{k,p}^i.$\\
Hence, we find that
\begin{align*}
\sum\limits_{i=1}^{+\infty} \parallel [A_k^i ]^{-1} \parallel_2 \le c_{k,q} \sum\limits_{i=1}^{+\infty} \theta_{k,p}^i = \dfrac{c_{k,p} \theta_{k,p}}{1-\theta_{k,p}},
\end{align*}
and the first point is proved.\\
The second one is analogous: $B_k$ has eigenvalues of modulus less than $1$, thus Lemma \ref{lemma:contrazione} provides two constants $c_{k,q} \in \mathbb{R}^+$ and $\theta_{k,q} \in (0,1)$ such that $\parallel B_k^{i} \parallel_2 \le c_{k,q} \theta_{k,q}^i,$ whenever $i \in \mathbb{N}$.\\
In conclusion,
\begin{align*}
\sum\limits_{i=0}^{+\infty} \parallel B_k^i  \parallel_2 \le c_{k,q} \sum\limits_{i=0}^{+\infty} \theta_{k,q}^i = \dfrac{c_{k,q}}{1-\theta_{k,q}},
\end{align*}
and the proposition is proved.
\end{proof}
%\ \\
%\vspace{-0.3cm}

\section{Proof of Theorem \ref{thrm:cd}}\label{section:proof_cd}
%\ \\

The theorem is equivalent to stating that, if $\widetilde{\delta}_N^{(1)}= \left[\widetilde{\lambda}_N^{(d+1)}\right]^{-1}$ represents the smallest eigenvalue of the normal matrix, then, under $C_d$, for every suitable pair $(x,k) \in X \times K$ and $N \in \mathbb{N}^*$ we have
\begin{align*}
\widetilde{\delta}_N^{(1)}(x,k) \le [M \sigma_k]^2(N+1).
\end{align*}\\
From Courant-Fischer Theorem, we find that \ $\widetilde{\delta}_N^{\, (1)} (x,k) = \min\limits_{\substack{\textbf{z} \in \mathbb{R}^{d+1} \\ \parallel \textbf{z} \parallel_2 =1}} \textbf{z}^T \widetilde{C}_N(x,k) \, \textbf{z}.$\\
Now, if we partition an arbitrary $\widetilde{\mathbf{v}} \in \mathbb{R}^{d+1}$ as
\begin{align*}
\widetilde{\mathbf{v}} = \begin{bmatrix}
\mathbf{v}\\
v_{d+1}
\end{bmatrix}, \ \ \ \text{with} \ \mathbf{v} \in \mathbb{R}^d \ \text{and} \ v_{d+1} \in \mathbb{R},
\end{align*}
by definition of $\widetilde{C}_N (x,k)$ and $\widetilde{F}_k^n (x)$, we get that
\begin{align*}
\begin{array}{l l}
\widetilde{\mathbf{v}}^T \widetilde{C}_N(x,k) \, \widetilde{\mathbf{v}} & = \sum\limits_{n=0}^N \parallel \widetilde{F}_k^n(x) \widetilde{\mathbf{v}} \parallel_2^2= \sum\limits_{n=0}^N \bigg| \bigg| \begin{bmatrix}
F_k^n (x) & \bigg| & \dfrac{\partial f_k^n}{\partial k}(x)
\end{bmatrix} \cdot \begin{bmatrix}
\mathbf{v}\\
v_{d+1}
\end{bmatrix} \bigg| \bigg|_2^2 =\\
\ \\
\ & = \sum\limits_{n=0}^N \bigg| \bigg| F_k^n (x) \cdot \mathbf{v} + v_{d+1} \cdot \dfrac{\partial f_k^n}{\partial k}(x) \bigg| \bigg|_2^2.
\end{array}
\end{align*}
Therefore, the crucial point is to find a vector $\widetilde{\mathbf{v}}$ independent on $N$ and such that
\begin{align*}
\bigg| \bigg| F_k^n (x) \cdot \mathbf{v} + v_{d+1} \cdot \dfrac{\partial f_k^n}{\partial k}(x) \bigg| \bigg|_2^2 \le [M \sigma_k]^2(N+1).
\end{align*}
For the next results, it is helpful to define an \textit{auxiliary map} as the function
\begin{center}
$\begin{array}{r c l l}
g: & X \times K & \rightarrow & X \times K\\
\ & \ & \ & \ \\
\ & (x,k) & \mapsto & (f_k(x), k).
\end{array}$
\end{center}
For every $n \in \mathbb{N}$, the $(d+1)\times (d+1)$ Jacobian matrix of its $n$-th iterate $g^n$ is well defined for every suitable pair $(x,k) \in X \times K$ and is given by
\begin{center}
$G^n (x,k):=\left[ \begin{array}{c | c}
F_k^n(x) & \dfrac{\partial f_k^n}{\partial k}(x)\\[1.4ex]
\hline\\[-1.3ex]
0 \dots 0 & 1
\end{array} \right]$.
\end{center}
Define the \textit{auxiliary normal matrix} as $C_N^g (x,k) \! := \! \sum\limits_{n=0}^N G^n (x,k)^T G^n (x,k)$, for every $N \! \in \! \mathbb{N}^*$.
\begin{lemma}
\label{matrice}
\ \\
For all $N \in \mathbb{N}^*$ and for every suitable pair $(x,k) \in X \times K$,
\begin{center}
$C_N^g (x,k) = \widetilde{C}_N (x,k) + \left[ \begin{array}{c | c}
\  & 0 \\
0_{d,d} & \vdots\\
\ & 0\\
\hline\\[-1.3ex]
0 \dots 0 & N+1
\end{array}\right]$.
\end{center}
\end{lemma}

\begin{proof}
The proof is straightforward from the definitions:
\begin{center}
$\begin{array}{l l}
C_N^g (x,k) & = \sum\limits_{n=0}^N G^n (x,k)^T G^n (x,k) = \sum\limits_{n=0}^N\left[ \begin{array}{c | c}
\  & 0\\
F_k^n(x)^T & \vdots \\
\ & 0\\
\hline\\[-1.3ex]
\dfrac{\partial f_k^n}{\partial k}(x)^T & 1
\end{array} \right] \left[ \begin{array}{c | c}
F_k^n(x) & \dfrac{\partial f_k^n}{\partial k}(x)\\[1.4ex]
\hline\\[-1.3ex]
0 \dots 0 & 1
\end{array} \right] =\\
\ & \ \\
\ & = \sum\limits_{n=0}^N \left[ \begin{array}{c | c}
F_k^n(x)^T F_k^n(x) & F_k^n(x)^T \dfrac{\partial f_k^n}{\partial k}(x) \\[1.4ex]
\hline\\[-1.3ex]
\dfrac{\partial f_k^n}{\partial k}(x)^T F_k^n(x) & \left\lvert \left\lvert \dfrac{\partial f_k^n}{\partial k}(x) \right\rvert \right\rvert_2^2 +1
\end{array} \right]=\\
\ & \ \\
\ & =\sum\limits_{n=0}^N \left[ \begin{array}{c | c}
F_k^n(x)^T F_k^n(x) & F_k^n(x)^T \dfrac{\partial f_k^n}{\partial k}(x) \\[1.4ex]
\hline\\[-1.3ex]
\dfrac{\partial f_k^n}{\partial k}(x)^T F_k^n(x) & \left\lvert \left\lvert \dfrac{\partial f_k^n}{\partial k}(x) \right\rvert \right\rvert_2^2
\end{array} \right] + \sum\limits_{n=0}^N \left[ \begin{array}{c | c}
\  & 0 \\
0 & \vdots\\
\ & 0\\
\hline\\[-1.3ex]
0 \dots 0 & 1
\end{array}\right] =\\
\ & \ \\
\ & = \widetilde{C}_N (x,k) + \left[ \begin{array}{c | c}
\  & 0 \\
0 & \vdots\\
\ & 0\\
\hline\\[-1.3ex]
0 \dots 0 & N+1
\end{array}\right].
\end{array}$
\end{center}
\end{proof}

\begin{lemma}
\ \\
For every suitable pair $(x,k) \in X \times K$ and $n \in \mathbb{N}$, the following equations are true:
\begin{equation}
\label{eqn:ricorsivo}
\begin{cases}
F_k^{n+1}(x) = F_k(f_k^n(x)) F_k^n(x);\\
\ \\
\dfrac{\partial f_k^{n+1}}{\partial k}(x) = F_k(f_k^n(x)) \dfrac{\partial f_k^n}{\partial k} (x) + \dfrac{\partial f_k}{\partial k} (f_k^n(x)).
\end{cases}
\end{equation}
\end{lemma}
\begin{proof} Both the equalities are implied by the chain rule.\\
The first one is straightforward from the definition $f_k^{n+1}(x) = f_k(f_k^n(x))$, while the second one is a little more obscure because of the notation of $f_k (x)$.\\
For the sake of completeness, we show both.\\
\\
Here, the auxiliary map $g$ is helpful: if we stress the fact that the couple $(x,k)$ varies, writing respectively $f(x,k)$, $F(x,k$), and $\widetilde{F}(x,k)$ instead of $f_k(x),F_k(x), \widetilde{F}_k(x)$, then the relation
\begin{align*}
f_k^{n+1}(x) = f_k(f_k^n (x))
\end{align*}
becomes
\begin{align*}
f_k^{n+1}(x) = f(f_k^n (x),k)= f(g^n (x,k)),
\end{align*}
and differentiating in $(x,k)$ employing the chain rule brings to
\begin{align*}
\begin{array}{l}
\widetilde{F}(g^n (x,k)) \, G^n (x,k) = \begin{bmatrix}
F(g^n(x,k)) & \bigg| & \dfrac{\partial f}{\partial k}(g^n(x,k))
\end{bmatrix} \, \left[ \begin{array}{c | c}
F^n(x,k) & \dfrac{\partial f^n}{\partial k}(x,k)\\[1.4ex]
\hline\\[-1.3ex]
0 \dots 0 & 1
\end{array} \right] =\\
\ \\
= \begin{bmatrix}
F(g^n(x,k))F^n(x,k) & \bigg| & F(g^n(x,k))\dfrac{\partial f^n}{\partial k}(x,k) + \dfrac{\partial f}{\partial k}(g^n(x,k))
\end{bmatrix}=\\
\ \\
 = \begin{bmatrix}
F(f_k^n(x),k)F^n(x,k) & \bigg| & F(f_k^n(x),k)\dfrac{\partial f^n}{\partial k}(x,k) + \dfrac{\partial f}{\partial k}(f_k^n(x),k)
\end{bmatrix}.
\end{array}
\end{align*}
\phantom{bla}\\
The first block of this $d \times (d+1)$ Jacobian matrix represents the partial derivatives of $f_k^{n+1}(x)$ with respect to $x$, that is $F_k^{n+1}(x)$.\\
The second block is related to the differentiation in $k$: $\dfrac{\partial f_k^{n+1}}{\partial k}(x)$.\\
Therefore, going back to the former notations, we find the equalities  (\ref{eqn:ricorsivo}).
\end{proof}

\begin{corollary}
\label{corollario:induttivo}
\ \\
For every suitable pair $(x , k) \in X \times K$ and $n \in \mathbb{N}^*$, we have that
\begin{equation}
\label{eqn:relinduttiva}
[F_k^n (x)]^{-1} \cdot \dfrac{\partial f_k^n}{\partial k}(x) = \sum\limits_{i=1}^n [F_k^i (x)]^{-1} \cdot \dfrac{\partial f_k}{\partial k}(f_k^{i-1}(x)).
\end{equation}
\end{corollary}

\begin{proof}
We prove the formula by induction on $n \ge 1$.\\
The step $n=1$ is rapidly checked.\\
\\
Let us suppose (\ref{eqn:relinduttiva}) is true for $n > 1$.\\
Using (\ref{eqn:ricorsivo}), we get:\\
\begin{align*}
\begin{array}{l}
[F_k^{n+1} (x)]^{-1} \cdot \dfrac{\partial f_k^{n+1}}{\partial k}(x) = [F_k^{n+1} (x)]^{-1} \cdot F_k(f_k^n(x)) \dfrac{\partial f_k^n}{\partial k} (x) + [F_k^{n+1} (x)]^{-1} \cdot \dfrac{\partial f_k}{\partial k} (f_k^n(x))=\\
\ \\
= [F_k^n(x)]^{-1} \cdot [F_k(f_k^n(x))]^{-1} \cdot F_k(f_k^n(x)) \cdot \dfrac{\partial f_k^n}{\partial k} (x) + [F_k^{n+1} (x)]^{-1} \cdot \dfrac{\partial f_k}{\partial k} (f_k^n(x)) = \\
\ \\
=[F_k^n(x)]^{-1}\cdot \dfrac{\partial f_k^n}{\partial k} (x) + [F_k^{n+1} (x)]^{-1} \cdot \dfrac{\partial f_k}{\partial k} (f_k^n(x)).
\end{array}
\end{align*}
The inductive hypothesis allows us to expand the first term in the last equality:
\begin{align*}
\begin{array}{l l}
[F_k^{n+1} (x)]^{-1} \cdot \dfrac{\partial f_k^{n+1}}{\partial k}(x) & = \sum\limits_{i=1}^n [F_k^i (x)]^{-1} \cdot \dfrac{\partial f_k}{\partial k}(f_k^{i-1}(x)) + [F_k^{n+1} (x)]^{-1} \cdot \dfrac{\partial f_k}{\partial k} (f_k^n(x))=\\
\ \\
\ & = \sum\limits_{i=1}^{n+1} [F_k^i (x)]^{-1} \cdot \dfrac{\partial f_k}{\partial k}(f_k^{i-1}(x)).
\end{array}
\end{align*}
hence (\ref{eqn:relinduttiva}) is proved.
\end{proof}
\phantom{bla}\\
The next lemma allows us to find a vector $\widetilde{\mathbf{v}}$ as above and such that \\
\begin{center}
$\bigg| \bigg| F_k^n (x) \cdot \mathbf{v} + v_{d+1} \cdot \dfrac{\partial f_k^n}{\partial k}(x) \bigg| \bigg|_2^2$
\end{center}
is bounded.

\begin{lemma}
\label{lemma:limite}
\ \\
Let us suppose that condition $C_d$ holds. Then, for a suitable pair $(x,k) \in X \times K$, the limit
\begin{align*}
L(x,k):=\lim\limits_{n \to +\infty} \left( [F_k^n (x)]^{-1} \cdot \dfrac{\partial f_k^n}{\partial k}(x) \right)
\end{align*}
exists.\\
Moreover, for every $n \in \mathbb{N}^*$
\begin{align*}
\left\lvert \left\lvert F_k^n(x) \cdot L(x,k) - \dfrac{\partial f_k^n}{\partial k}(x) \right\rvert \right\rvert_2 \le M \sigma_k.
\end{align*}
\end{lemma}

\begin{proof} Using Corollary \ref{corollario:induttivo}, we note that, for a suitable pair $(x,k) \in X \times K$, the existence of the limit $L(x,k)$ is equivalent to the convergence of the series \ $\sum\limits_{i=1}^{+\infty} [F_k^i (x)]^{-1} \cdot \dfrac{\partial f_k}{\partial k}(f_k^{i-1}(x)),$\\

\vspace{0.1cm}which we obtain from the convergence of \ $\sum\limits_{i=1}^{+\infty} \left\lvert \left\lvert [F_k^i (x)]^{-1} \cdot \dfrac{\partial f_k}{\partial k}(f_k^{i-1}(x)) \right\rvert \right\rvert_2$.\\

\vspace{0.1cm}Knowing that, for every $i \in \mathbb{N}^*$, we have $\bigg| \bigg| \dfrac{\partial f_k}{\partial k}(f_k^{i-1}(x)) \bigg| \bigg|_2 \le M,$ we find
\vspace{0.2cm}
\begin{align*}
\begin{array}{l l}
\sum\limits_{i=1}^{+\infty} \left\lvert \left\lvert [F_k^i (x)]^{-1} \cdot \dfrac{\partial f_k}{\partial k}(f_k^{i-1}(x)) \right\rvert \right\rvert_2 & \le \sum\limits_{i=1}^{+\infty} \parallel [F_k^i (x)]^{-1} \parallel_2 \cdot \left\lvert \left\lvert \dfrac{\partial f_k}{\partial k}(f_k^{i-1}(x)) \right\rvert \right\rvert_2 \le \\
\ \\
\ & \le M \sum\limits_{i=1}^{+\infty} \parallel [F_k^i (x)]^{-1} \parallel_2,
\end{array}
\end{align*}
and the convergence follows by the condition $C_d$.\\
\\
Now we turn to the second part of the lemma.\\
Thanks to equation (\ref{eqn:relinduttiva}) and the first part of the proof, we have:\\
\begin{align*}
\begin{array}{l}
F_k^n(x) \cdot L(x,k) - \dfrac{\partial f_k^n}{\partial k}(x) = F_k^n(x) \left( L(x,k) - [F_k^n(x)]^{-1} \cdot \dfrac{\partial f_k^n}{\partial k}(x) \right) = \\
\ \\
=F_k^n(x) \left( \sum\limits_{i=1}^{+\infty} [F_k^i (x)]^{-1} \cdot \dfrac{\partial f_k}{\partial k}(f_k^{i-1}(x)) - \sum\limits_{i=1}^{n} [F_k^i (x)]^{-1} \cdot \dfrac{\partial f_k}{\partial k}(f_k^{i-1}(x)) \right)=\\
\ \\
= F_k^n(x) \cdot \sum\limits_{i=n+1}^{+\infty} [F_k^i (x)]^{-1} \cdot \dfrac{\partial f_k}{\partial k}(f_k^{i-1}(x)).
\end{array}
\end{align*}
\phantom{bla}\\
Using the chain rule, we note that, for every $i \ge n+1$,
\begin{align*}
\begin{array}{l l}
F_k^n(x) [F_k^i(x)]^{-1} & = F_k^n(x) [F_k^{i-n}(f_k^n(x)) F_k^n(x)]^{-1} = F_k^n(x) [F_k^n(x)]^{-1} [F_k^{i-n}(f_k^n(x))]^{-1}=\\
\ \\
\ & =[F_k^{i-n}(f_k^n(x))]^{-1}.
\end{array}
\end{align*}
Thus
\begin{align*}
F_k^n(x) \cdot L(x,k) - \dfrac{\partial f_k^n}{\partial k}(x) = \sum\limits_{i=n+1}^{+\infty} [F_k^{i-n} (f_k^n(x))]^{-1} \cdot \dfrac{\partial f_k}{\partial k}(f_k^{i-1}(x)),
\end{align*}
and taking $j=i-n$ as summation index, we get
\begin{align*}
F_k^n(x) \cdot L(x,k) - \dfrac{\partial f_k^n}{\partial k}(x) = \sum\limits_{j=1}^{+\infty} [F_k^{j} (f_k^n(x))]^{-1} \cdot \dfrac{\partial f_k}{\partial k}(f_k^{n+j-1}(x)).
\end{align*}
In conclusion, we can estimate the $2$-norm from above:
\begin{align*}
\begin{array}{l l}
\left\lvert \left\lvert F_k^n(x) \cdot L(x,k) - \dfrac{\partial f_k^n}{\partial k}(x) \right\rvert \right\rvert_2 & \le \sum\limits_{j=1}^{+\infty} \left\lvert \left\lvert [F_k^{j} (f_k^n(x))]^{-1} \cdot \dfrac{\partial f_k}{\partial k}(f_k^{n+j-1}(x)) \right\rvert \right\rvert_2 \le\\
\ \\
\ & \le M \sum\limits_{j=1}^{+\infty} \left\lvert \left\lvert [F_k^{j} (f_k^n(x))]^{-1} \right\rvert \right\rvert_2 \le M \sigma_k,
\end{array}
\end{align*}
and the second part of the lemma is proven.
\end{proof}

We are now ready to solve the problem of finding a vector $\widetilde{\textbf{v}} \in \mathbb{R}^{d+1}$ with $\parallel \widetilde{\textbf{v}} \parallel_2 = 1$ such that $\widetilde{\textbf{v}}^T \widetilde{C}_N (x,k) \, \widetilde{\textbf{v}} \le [M \sigma_k]^2 (N+1)$: fixed a suitable couple $(x,k)$, we define
\begin{align*}
\mathbf{v}(x,k) := \dfrac{L(x,k)}{\sqrt{\parallel L(x,k) \parallel_2^2 + 1}} \in \mathbb{R}^d \ \ \ \ \text{and} \ \ \ \ v_{d+1}:= \dfrac{-1}{\sqrt{\parallel L(x,k) \parallel_2^2 + 1}} \in \mathbb{R},
\end{align*}
where $L(x,k) \in \mathbb{R}^d$ is the limit defined in Lemma \ref{lemma:limite}, and assemble them into
\begin{align*}
\widetilde{\mathbf{v}}(x,k) := \begin{bmatrix}
\mathbf{v}(x,k) \\
v_{d+1}(x,k)
\end{bmatrix},
\end{align*}
which, by construction, has norm $1$. Thus,
\begin{align*}
\begin{array}{l}
\widetilde{\mathbf{v}}(x,k)^T \widetilde{C}_N(x,k) \widetilde{\mathbf{v}}(x,k) = \sum\limits_{n=0}^N \bigg| \bigg| F_k^n (x) \cdot \mathbf{v}(x,k) + v_{d+1}(x,k) \cdot \dfrac{\partial f_k^n}{\partial k}(x) \bigg| \bigg|_2^2 = \\
\ \\
= \dfrac{1}{\parallel L(x,k) \parallel_2^2 + 1} \cdot \sum\limits_{n=0}^N \bigg| \bigg| F_k^n(x) L(x,k) - \dfrac{\partial f_k^n}{\partial k}(x) \bigg| \bigg|_2^2 \le \dfrac{[M \sigma_k ]^2}{\parallel L(x,k) \parallel_2^2 + 1} (N+1),
\end{array}
\end{align*}
where the last step follows from Lemma \ref{lemma:limite}.\\
Hence, we can conclude that, by Courant-Fischer Theorem,
\begin{align*}
\begin{array}{l l}
\widetilde{\delta}_N^{\, (1)} (x,k) & = \min\limits_{\substack{\textbf{z} \in \mathbb{R}^{d+1} \\ \parallel \textbf{z} \parallel_2 =1}} \textbf{z}^T \widetilde{C}_N(x,k) \, \textbf{z} \le \widetilde{\textbf{v}}^T \widetilde{C}_N(x,k) \, \widetilde{\textbf{v}} \le \\
\\
\ & \le \dfrac{[M \sigma_k ]^2}{\parallel L(x,k) \parallel_2^2 + 1} (N+1) \le [M \sigma_k ]^2 (N+1),
\end{array}
\end{align*}
being $\parallel L(x,k) \parallel_2^2 + 1 \ge 1$.\\
\\
Now, the proof easily follows recalling that $\widetilde{\lambda}_N^{(d+1)}(x,k) = \left[ \widetilde{\delta}_N^{\, (1)} (x,k) \right]^{-1}.$\\

\section{Proof of Theorem \ref{thrm:hd}}\label{section:proofthrm2}
%\ \\

By assumption, for every suitable pair $(x, k) \in X \times K$ and for every $n \in \mathbb{N}^*$, we have
\begin{align*}
\begin{array}{l}
F_k^n(x)= F_k(f_k^{n-1}(x)) F_k(f_k^{n-2}(x)) \dots F_k(f_k(x)) F_k (x) = \\
\\
= V_k V_k^{-1}F_k(f_k^{n-1}(x))V_k V_k^{-1}F_k(f_k^{n-2}(x))  \dots V_k V_k^{-1} F_k(f_k(x)) V_k V_k^{-1} F_k (x) V_k V_k^{-1} =\\
\\
= V_k D_k (f_k^{n-1}(x)) D_k (f_k^{n-2}(x)) \dots D_k(f_k(x)) D_k (x) V_k^{-1}=\\
\\
= V_k D_k^n(x) V_k^{-1},
\end{array}
\end{align*}
\phantom{bla}\\
where 
\begin{align*}
D_k^n(x) & := D_k (f_k^{n-1}(x)) D_k (f_k^{n-2}(x)) \dots D_k(f_k(x)) D_k (x) = \left[ \begin{array}{c | c}
A_k^n(x) & 0_{p,q}\\[1.4ex]
\hline\\[-1.3ex]
0_{q,p} & B_k^n(x)
\end{array} \right].
\end{align*}

From the second equation in (\ref{eqn:ricorsivo}) we have that\\
\begin{align*}
\dfrac{\partial f_k^{n+1}}{\partial k}(x) = F_k(f_k^n(x)) \dfrac{\partial f_k^n}{\partial k} (x) + \dfrac{\partial f_k}{\partial k} (f_k^n(x)),
\end{align*}

for a suitable pair $(x,k) \in X \times K$, hence we can deduce another representation for the derivatives with respect to $k$.

\begin{lemma}
\label{lemma:altrarelazione}
\ \\
For every suitable $(x, k )  \in X \times K$ and for every $n \in \mathbb{N}^*$, we have that
\begin{align*}
\dfrac{\partial f_k^{n}}{\partial k}(x) = \sum\limits_{i=0}^{n-1} F_k^i (f_k^{n-i}(x)) \cdot \dfrac{\partial f_k}{\partial k}(f_k^{n-1-i}(x)).
\end{align*}
\end{lemma}

\begin{proof}
We can show the formula by induction on $n$.\\
\\
When $n=1$ this  fact is immediate.\\
If $n>1$, then, employing the second of equations (\ref{eqn:ricorsivo}) and the inductive hypothesis, we get
\begin{align*}
\begin{array}{l l}
\dfrac{\partial f_k^{n}}{\partial k}(x) & = F_k(f_k^{n-1}(x))\dfrac{\partial f_k^{n-1}}{\partial k}(x) + \dfrac{\partial f_k}{\partial k}(f_k^{n-1}(x)) = \\
\ \\
\ &= F_k(f_k^{n-1}(x)) \cdot \sum\limits_{i=0}^{n-2} F_k^i (f_k^{n-1-i}(x)) \cdot \dfrac{\partial f_k}{\partial k}(f_k^{n-2-i}(x)) + \dfrac{\partial f_k}{\partial k}(f_k^{n-1}(x))= \\
\ \\
\ & = \sum\limits_{i=0}^{n-2} F_k(f_k^{n-1}(x)) F_k^i (f_k^{n-1-i}(x)) \cdot \dfrac{\partial f_k}{\partial k}(f_k^{n-2-i}(x)) + \dfrac{\partial f_k}{\partial k}(f_k^{n-1}(x)).
\end{array}
\end{align*}
Noting that
\begin{align*}
\begin{array}{l l}
F_k(f_k^{n-1}(x)) F_k^i (f_k^{n-1-i}(x)) & = F_k(f_k^{n-1}(x)) F_k(f_k^{i-1}(f_k^{n-1-i}(x))) \dots F_k(f_k^{n-1-i}(x)) = \\
\\
\ & = F_k ( f_k^i (f_k^{n-1-i}(x))) F_k(f_k^{i-1}(f_k^{n-1-i}(x))) \dots F_k(f_k^{n-1-i}(x)) = \\
\\
\ & = F_k^{i+1} (f_k^{n-1-i} (x)),
\end{array}
\end{align*}
we have
\begin{align*}
\begin{array}{l l}
\dfrac{\partial f_k^{n}}{\partial k}(x) & = \sum\limits_{i=0}^{n-2} F_k^{i+1}(f_k^{n-1-i}(x)) \cdot \dfrac{\partial f_k}{\partial k}(f_k^{n-2-i}(x)) + I_{d \times d}\dfrac{\partial f_k}{\partial k}(f_k^{n-1}(x))=\\
\ \\
\ & = \sum\limits_{i=1}^{n-1} F_k^{i}(f_k^{n-i}(x)) \cdot \dfrac{\partial f_k}{\partial k}(f_k^{n-1-i}(x)) + F_k^0(f_k^n(x))\dfrac{\partial f_k }{\partial k}(f_k^{n-1}(x))=\\
\ \\
\ & = \sum\limits_{i=0}^{n-1} F_k^i (f_k^{n-i}(x)) \cdot \dfrac{\partial f_k}{\partial k}(f_k^{n-1-i}(x)),
\end{array}
\end{align*}
and the lemma is proved.
\end{proof}
Given a suitable $(x, k) \in X \times K$, and $n \in \mathbb{N}^*$, we may express $\dfrac{\partial f_k^{n}}{\partial k}(x)$ in terms of the matrices $D_k^i(x)$, with $i \in \lbrace 0 , \dots , n-1 \rbrace$:
\begin{align*}
\begin{array}{l l}
\dfrac{\partial f_k^{n}}{\partial k}(x) & = \sum\limits_{i=0}^{n-1} F_k^i (f_k^{n-i}(x)) \cdot \dfrac{\partial f_k}{\partial k}(f_k^{n-1-i}(x)) = \sum\limits_{i=0}^{n-1} V_k D_k^i (f_k^{n-i}(x)) V_k^{-1} \cdot \dfrac{\partial f_k}{\partial k}(f_k^{n-1-i}(x))=\\
\ \\
\ & =V_k \sum\limits_{i=0}^{n-1} \left[ \begin{array}{c | c}
A_k^i(f_k^{n-i}(x)) & 0_{p,q}\\[1.4ex]
\hline\\[-1.3ex]
0_{q,p} & B_k^i(f_k^{n-i}(x))
\end{array} \right] \cdot V_k^{-1} \cdot \dfrac{\partial f_k}{\partial k}(f_k^{n-1-i}(x)).
\end{array}
\end{align*}

Now, denoting for every suitable $\overline{x} \in X$
\begin{align*}
\begin{bmatrix}
\boldsymbol{\omega}_k^1(\overline{x})\\
\boldsymbol{\omega}_k^2(\overline{x})
\end{bmatrix}:=V_k^{-1} \cdot \dfrac{\partial f_k}{\partial k}(\overline{x}) ,
\end{align*}
where $\boldsymbol{\omega}_k^1(\overline{x}) \in \mathbb{R}^p$ and $\boldsymbol{\omega}_k^2(\overline{x}) \in \mathbb{R}^q$, we obtain
\begin{align*}
\dfrac{\partial f_k^{n}}{\partial k}(x) =  V_k  \begin{bmatrix}
\sum\limits_{i=0}^{n-1}A_k^i(f_k^{n-i}(x)) \cdot \boldsymbol{\omega}_k^1(f_k^{n-1-i}(x))\\[1.4ex]
\ \\
\sum\limits_{i=0}^{n-1}B_k^i(f_k^{n-i}(x)) \cdot \boldsymbol{\omega}_k^2(f_k^{n-1-i}(x))
\end{bmatrix}.
\end{align*}

Moreover, we note that for $j=1,2$,
\begin{equation}
\label{eqn:stima.omega}
\parallel \boldsymbol{\omega}_k^j(\overline{x})\parallel_2 \, \le \, \parallel V_k^{-1} \dfrac{\partial f_k}{\partial k}(\overline{x}) \parallel_2 \, \le \, M \parallel V_k^{-1} \parallel_2.
\end{equation}
Before stating the main result of this section, we need some facts similar to those of Lemma \ref{lemma:limite} (the proof is analogous).

\begin{lemma}
\label{lemma:limite3}
\ \\
If condition $H_d$ is satisfied, then, for every suitable pair $(x,k) \in X \times K$, the limit
\begin{align*}
L(x,k):= \lim\limits_{n \to +\infty} \left([A_k^n(x)]^{-1} \sum\limits_{i=0}^{n-1} A_k^i (f_k^{n-i}(x))\cdot \boldsymbol{\omega}_k^1 (f_k^{n-1-i}(x)) \right)
\end{align*}
exists.\\
\\
Moreover, for every $n \in \mathbb{N}^*$
\begin{align*}
\bigg| \bigg| A_k^n(x) \cdot L(x,k) - \sum\limits_{i=0}^{n-1}A_k^i(f_k^{n-i}(x)) \cdot \boldsymbol{\omega}_k^1 (f_k^{n-1-i}(x)) \bigg| \bigg|_2 \, \le \, M \parallel V_k^{-1} \parallel_2 \cdot \alpha_k.
\end{align*}
\end{lemma}

\begin{proof}
Let us fix a suitable couple $(x,k) \in X \times K$, $n \in \mathbb{N}^*$ and $i \in \lbrace 0 , \dots , n-1 \rbrace$. We note that
\begin{align*}
\begin{array}{l l}
[A_k^n (x) ]^{-1} A_k^i (f_k^{n-i} (x)) & =  [A_k (f_k^{n-1}(x)) \dots A_k (x)]^{-1} A_k(f_k^{i-1}(f_k^{n-i}(x))) \dots A_k (f_k^{n-i}(x))=\\
\\
\ & = [A_k (x)]^{-1} \dots [A_k (f_k^{n-1}(x))]^{-1}  A_k (f_k^{n-1}(x)) \dots A_k (f_k^{n-i}(x))=\\
\\
\ & = [A_k (x)]^{-1} \dots [A_k (f_k^{n-1-i}(x))]^{-1}=\\
\\
\ & =[ A_k^{n-i} (x) ]^{-1}.
\end{array}
\end{align*}
Hence,
\begin{align*}
\begin{array}{l l}
[A_k^n(x)]^{-1} \sum\limits_{i=0}^{n-1} A_k^i (f_k^{n-i}(x))\cdot \boldsymbol{\omega}_k^1 (f_k^{n-1-i}(x)) & = \sum\limits_{i=0}^{n-1} [ A_k^{n-i} (x) ]^{-1} \cdot \boldsymbol{\omega}_k^1 (f_k^{n-1-i}(x)) = \\
\\
\ & = \sum\limits_{i=1}^{n} [ A_k^{i} (x) ]^{-1} \cdot \boldsymbol{\omega}_k^1 (f_k^{i-1}(x)),
\end{array}
\end{align*}
and thus the existence of the limit $L(x,k)$ is equivalent to the convergence of the series
\begin{align*}
\sum\limits_{i=1}^{+\infty} [ A_k^{i} (x) ]^{-1} \cdot \boldsymbol{\omega}_k^1 (f_k^{i-1}(x)).
\end{align*}
In order to achieve this, we show that the series of the norms
\begin{align*}
\sum\limits_{i=1}^{+\infty} \left\lvert \left\lvert [ A_k^{i} (x) ]^{-1} \cdot \boldsymbol{\omega}_k^1 (f_k^{i-1}(x)) \right\rvert \right\rvert_2
\end{align*}
converges.\\
\\
From (\ref{eqn:stima.omega}), we deduce that, whenever $i \in \mathbb{N}^*$,
\begin{align*}
\bigg| \bigg| \boldsymbol{\omega}_k^1 (f_k^{i-1}(x))\bigg| \bigg|_2 \le M \parallel V_k^{-1} \parallel_2,
\end{align*}
so condition $H_d$ implies
\begin{align*}
\sum\limits_{i=1}^{+\infty} \left\lvert \left\lvert [ A_k^{i} (x) ]^{-1} \cdot \boldsymbol{\omega}_k^1 (f_k^{i-1}(x)) \right\rvert \right\rvert_2 \le  M \parallel V_k^{-1} \parallel_2  \sum\limits_{i=1}^{+\infty} \left\lvert \left\lvert [ \, A_k^{i} (x) \, ]^{-1} \right\rvert \right\rvert_2 < + \infty,
\end{align*}
and the first part of the lemma is proved.\\
The second one is also analogous to Lemma \ref{lemma:limite}.\\
In particular, we can write:
\begin{align*}
\begin{array}{l}
A_k^n(x) \cdot L(x,k) - \sum\limits_{i=0}^{n-1} A_k^i (f_k^{n-i}(x))\cdot \boldsymbol{\omega}_k^1 (f_k^{n-1-i}(x)) =\\
\\
 = A_k^n(x) \left( L(x,k) - [A_k^n(x)]^{-1} \cdot \sum\limits_{i=0}^{n-1} A_k^i (f_k^{n-i}(x))\cdot \boldsymbol{\omega}_k^1 (f_k^{n-1-i}(x)) \right) = \\
\ \\
=A_k^n(x) \left( \sum\limits_{i=1}^{+\infty} [ A_k^{i} (x) ]^{-1} \cdot \boldsymbol{\omega}_k^1 (f_k^{i-1}(x)) - \sum\limits_{i=1}^{n} [ A_k^{i} (x) ]^{-1} \cdot \boldsymbol{\omega}_k^1 (f_k^{i-1}(x)) \right)=\\
\ \\
= A_k^n(x) \cdot \sum\limits_{i=n+1}^{+\infty} [A_k^i (x)]^{-1} \cdot \boldsymbol{\omega}_k^1 (f_k^{i-1}(x)),
\end{array}
\end{align*}

where in the second-last passage we used again the fact that
\begin{align*}
[A_k^n(x)]^{-1} \sum\limits_{i=0}^{n-1} A_k^i (f_k^{n-i}(x))\cdot \boldsymbol{\omega}_k^1 (f_k^{n-1-i}(x))  = \sum\limits_{i=1}^{n} [ A_k^{i} (x) ]^{-1} \cdot \boldsymbol{\omega}_k^1 (f_k^{i-1}(x)),
\end{align*}
\phantom{bla}\\
Now, for every $i \ge n+1$,
\begin{align*}
\begin{array}{l l}
A_k^n(x) [A_k^i(x)]^{-1} & = A_k^n(x) [A_k^{i-n}(f_k^n(x)) A_k^n(x)]^{-1} = A_k^n(x) [A_k^n(x)]^{-1} [A_k^{i-n}(f_k^n(x))]^{-1} = \\
\\
\ & =[A_k^{i-n}(f_k^n(x))]^{-1}.
\end{array}
\end{align*}
Thus
\begin{align*}
\begin{array}{l l}
A_k^n(x) \cdot L(x,k) - \sum\limits_{i=0}^{n-1} A_k^i (f_k^{n-i}(x))\cdot \boldsymbol{\omega}_k^1 (f_k^{n-1-i}(x)) & = \sum\limits_{i=n+1}^{+\infty} [A_k^{i-n} (f_k^n(x))]^{-1} \cdot \boldsymbol{\omega}_k^1 (f_k^{i-1}(x)) = \\
\\
\ & = \sum\limits_{j=1}^{+\infty} [A_k^{j} (f_k^n(x))]^{-1} \cdot \boldsymbol{\omega}_k^1 (f_k^{n+j-1}(x)).
\end{array}
\end{align*}
Therefore, we can conclude:
\begin{align*}
\begin{array}{l l}
\left\lvert \left\lvert A_k^n(x) \cdot L(x,k) - \sum\limits_{i=0}^{n-1} A_k^i (f_k^{n-i}(x))\cdot \boldsymbol{\omega}_k^1 (f_k^{n-1-i}(x)) \right\rvert \right\rvert_2 & \le \sum\limits_{j=1}^{+\infty} \left\lvert \left\lvert [A_k^{j} (f_k^n(x))]^{-1} \cdot \boldsymbol{\omega}_k^1 (f_k^{n+j-1}(x)) \right\rvert \right\rvert_2 \le\\
\ \\
\ & \le M \parallel V_k^{-1} \parallel_2 \sum\limits_{j=1}^{+\infty} \left\lvert \left\lvert [A_k^{j} (f_k^n(x))]^{-1} \right\rvert \right\rvert_2\\
\ \\
\ & \le M \parallel V_k^{-1} \parallel_2 \alpha_k.
\end{array}
\end{align*}
\end{proof}
For what comes next, it is convenient to define the matrix
\begin{align*}
V_k^{(p)} := V_k \cdot \begin{bmatrix}
I_{p \times p}\\
0_{q \times p}
\end{bmatrix} \in \mathbb{R}^{d \times p},
\end{align*}
that is, the matrix consisting of the first $p$ columns of $V_k$.\\
\\
We know that $\widetilde{\lambda}_N^{(d+1)}(x,k)$ is the reciprocal of the smallest eigenvalue of the normal matrix, which by the Courant-Fischer Theorem can be expressed as
\begin{align*}
\widetilde{\delta}_N^{\, (1)}(x,k) = \min\limits_{\substack{\textbf{z} \in \mathbb{R}^{d+1} \\ \parallel \textbf{z} \parallel_2 =1}} \textbf{z}^T \widetilde{C}_N(x,k) \, \textbf{z}.
\end{align*}
Therefore, we look for an estimate for $\widetilde{\delta}_N^{\, (1)}(x,k)$ from above.\\
\\
Just like in the previous section, we know that if we write an arbitrary $\widetilde{\mathbf{v}} \in \mathbb{R}^{d+1}$ as
\begin{align*}
\widetilde{\mathbf{v}} = \begin{bmatrix}
\mathbf{v}\\
v_{d+1}
\end{bmatrix}, \ \ \ \text{with} \ \mathbf{v} \in \mathbb{R}^d \ \text{and} \ v_{d+1} \in \mathbb{R},
\end{align*}
we get that
\begin{align*}
\widetilde{\mathbf{v}}^T \widetilde{C}_N(x,k) \, \widetilde{\mathbf{v}} =  \sum\limits_{n=0}^N \bigg| \bigg| F_k^n (x) \cdot \mathbf{v} + v_{d+1} \cdot \dfrac{\partial f_k^n}{\partial k}(x) \bigg| \bigg|_2^2.
\end{align*}
We now look for a suitable $\widetilde{\mathbf{v}}$ that leads to the proof.\\
\\
Let $L(x,k) \in \mathbb{R}^p$ be the limit as in Lemma \ref{lemma:limite3}. We define
\begin{align*}
\mathbf{v}(x,k):= \dfrac{V_k^{(p)} \cdot L(x,k)}{\sqrt{\parallel V_k^{(p)} \cdot L(x,k) \parallel_2^2+1}} = \dfrac{V_k \cdot \begin{bmatrix}
L(x,k)\\
0_{q,1}
\end{bmatrix}}{\sqrt{\parallel V_k^{(p)} \cdot L(x,k) \parallel_2^2+1}} \in \mathbb{R}^d
\end{align*}
and
\begin{align*}
v_{d+1}(x,k):= -\dfrac{1}{\sqrt{\parallel V_k^{(p)} \cdot L(x,k) \parallel_2^2+1}},
\end{align*}
then we assemble them into
\begin{align*}
\widetilde{\mathbf{v}}(x,k) := \begin{bmatrix}
\mathbf{v}(x,k) \\
v_{d+1}(x,k)
\end{bmatrix},
\end{align*}
which, by construction, has norm $1$.\\
\\
Thus,
\begin{align*}
\widetilde{\mathbf{v}}(x,k)^T \widetilde{C}_N(x,k) \widetilde{\mathbf{v}}(x,k) = \sum\limits_{n=0}^N \bigg| \bigg| F_k^n (x) \cdot \mathbf{v}(x,k) + v_{d+1}(x,k) \cdot \dfrac{\partial f_k^n}{\partial k}(x) \bigg| \bigg|_2^2 = \\
\ \\
= \dfrac{1}{\parallel V_k^{(p)} \cdot L(x,k) \parallel_2^2+1} \cdot \sum\limits_{n=0}^N \bigg| \bigg| F_k^n(x) \cdot V_k \cdot \begin{bmatrix}
L(x,k)\\
0_{q,1}
\end{bmatrix} - \dfrac{\partial f_k^n}{\partial k}(x) \bigg| \bigg|_2^2.
\end{align*}

Let us focus, for every $n = 0, \dots , N$, on the term
\begin{align*}
F_k^n(x) \cdot V_k \cdot \begin{bmatrix}
L(x,k)\\
0_{q,1}
\end{bmatrix} - \dfrac{\partial f_k^n}{\partial k}(x).
\end{align*}
We recall that
\begin{align*}
F_k^n(x) = V_k \cdot \left[ \begin{array}{c | c}
A_k^n(x) & 0_{p,q}\\[1.4ex]
\hline\\[-1.3ex]
0_{q,p} & B_k^n(x)
\end{array} \right] \cdot V_k^{-1}
\end{align*}
and
\begin{align*}
\dfrac{\partial f_k^{n}}{\partial k}(x) =  V_k \cdot \begin{bmatrix}
\sum\limits_{i=0}^{n-1}A_k^i(f_k^{n-i}(x)) \cdot \boldsymbol{\omega}_k^1(f_k^{n-1-i}(x))\\[1.4ex]
\ \\
\sum\limits_{i=0}^{n-1}B_k^i(f_k^{n-i}(x)) \cdot \boldsymbol{\omega}_k^2(f_k^{n-1-i}(x))
\end{bmatrix},
\end{align*}
so
\begin{align*}
\begin{array}{l}
F_k^n(x) \cdot V_k \cdot \begin{bmatrix}
L(x,k)\\
0_{q,1}
\end{bmatrix} - \dfrac{\partial f_k^n}{\partial k}(x) = \\
\ \\
= V_k \cdot \left[ \begin{array}{c | c}
A_k^n(x) & 0_{p,q}\\[1.4ex]
\hline\\[-1.3ex]
0_{q,p} & B_k^n(x)
\end{array} \right] \cdot V_k^{-1} \cdot V_k \cdot \begin{bmatrix}
L(x,k)\\
0_{q,1}
\end{bmatrix} - V_k \cdot \begin{bmatrix}
\sum\limits_{i=0}^{n-1}A_k^i(f_k^{n-i}(x)) \cdot \boldsymbol{\omega}_k^1(f_k^{n-1-i}(x))\\[1.4ex]
\ \\
\sum\limits_{i=0}^{n-1}B_k^i(f_k^{n-i}(x)) \cdot \boldsymbol{\omega}_k^2(f_k^{n-1-i}(x))
\end{bmatrix}.
\end{array}
\end{align*}
\phantom{bla}\\
Simplifying the term $V_k^{-1} \cdot V_k$ and factoring out $V_k$ on the left, we get\\
\begin{align*}
F_k^n(x) \cdot V_k \cdot \begin{bmatrix}
L(x,k)\\
0_{q,1}
\end{bmatrix} - \dfrac{\partial f_k^n}{\partial k}(x) = V_k \cdot \begin{bmatrix}
A_k^n(x) L(x,k) - \sum\limits_{i=0}^{n-1}A_k^i (f_k^{n-i}(x)) \cdot \boldsymbol{\omega}_k^1 (f_k^{n-1-i}(x))\\
\sum\limits_{i=0}^{n-1} B_k^i (f_k^{n-i}(x)) \cdot \boldsymbol{\omega}_k^2 (f_k^{n-1-i}(x))
\end{bmatrix}.
\end{align*}
\phantom{bla}\\
Now, the squared $2-$norm of this vector is not bigger than\\
\begin{align*}
\parallel V_k \parallel_2^2 \bigg( \, \bigg| \bigg| A_k^n(x) L(x,k)\ \  -&\! \!  \sum\limits_{i=0}^{n-1}A_k^i (f_k^{n-i}(x)) \cdot \boldsymbol{\omega}_k^1 (f_k^{n-1-i}(x))\bigg| \bigg|_2^2 + \bigg| \bigg| \sum\limits_{i=0}^{n-1} B_k^i (f_k^{n-i}(x)) \cdot \boldsymbol{\omega}_k^2 (f_k^{n-1-i}(x)) \bigg| \bigg|_2^2 \, \bigg).
\end{align*}
Lemma \ref{lemma:limite3} guarantees that\\
\begin{align*}
\bigg| \bigg| A_k^n(x) L(x,k) - \sum\limits_{i=0}^{n-1}A_k^i (f_k^{n-i}(x)) \cdot \boldsymbol{\omega}_k^1 (f_k^{n-1-i}(x))\bigg| \bigg|_2^2 \, \le \, M^2 \parallel V_k^{-1} \parallel_2^2 \alpha_k^2,
\end{align*}
while condition $H_d$ (the part regarding $B_k(x)$) and the estimate
\begin{align*}
\parallel \boldsymbol{\omega}_k^2(\overline{x})\parallel_2 \, \le \, M \parallel V_k^{-1} \parallel_2 \ \text{for every suitable} \ \overline{x} \in X
\end{align*}
imply that
\begin{align*}
\begin{array}{l l}
\bigg| \bigg| \sum\limits_{i=0}^{n-1} B_k^i (f_k^{n-i}(x)) \cdot \boldsymbol{\omega}_k^2 (f_k^{n-1-i}(x)) \bigg| \bigg|_2^2 \, & \le \, M^2 \parallel V_k^{-1} \parallel_2^2 \left[ \sum\limits_{i=0}^{n-1} \parallel B_k^i (f_k^{n-i}(x)) \parallel_2 \right]^2 \le \\
\\
\ & \le M^2 \parallel V_k^{-1} \parallel_2^2 \beta_k^2.
\end{array}
\end{align*}
Collecting all this facts, we find that
\begin{align*}
\begin{array}{l}
\widetilde{\mathbf{v}}(x,k)^T \widetilde{C}_N(x,k) \widetilde{\mathbf{v}}(x,k) \, \le \\
\\
 \le \dfrac{1}{\parallel V_k^{(p)} \cdot L(x,k) \parallel_2^2 +1} \cdot \sum\limits_{n=0}^N  \parallel V_k \parallel_2^2 \left( M^2 \parallel V_k^{-1} \parallel_2^2 \alpha_k^2 + M^2 \parallel V_k^{-1} \parallel_2^2 \beta_k^2 \right)=\\
\ \\
=\dfrac{\left( M \parallel V_k \parallel_2 \parallel V_k^{-1} \parallel_2 \right)^2 (\alpha_k^2 + \beta_k^2)}{\parallel V_k^{(p)} \cdot L(x,k) \parallel_2^2 +1} \cdot (N+1),
\end{array}
\end{align*}

which rapidly leads to the proof, being
\begin{align*}
\widetilde{\delta}_N^{\, (1)}(x,k) = \min\limits_{\substack{\textbf{z} \in \mathbb{R}^{d+1} \\ \parallel \textbf{z} \parallel_2 =1}} \textbf{z}^T \widetilde{C}_N(x,k) \, \textbf{z} \le \widetilde{\mathbf{v}}(x,k)^T \widetilde{C}_N(x,k) \widetilde{\mathbf{v}}(x,k).
\end{align*}
and
\begin{align*}
\widetilde{\lambda}_N^{(d+1)}(x,k) = \left[\widetilde{\delta}_N^{\, (1)}(x,k)\right]^{-1}.
\end{align*}
%\ \\

\section{Intermittent maps}\label{section:intermittent}
%\ \\

In some cases, conditions $C_d$ and $H_d$ may turn out to be too strict. Intermittent maps of the unit interval are an example of transformations not satisfying $C_1$ (and neither $H_1$ as a consequence).\\
More specifically, we consider a family $\lbrace f_k: [0,1] \rightarrow [0,1] \rbrace_{k \in K}$ defined piecewisely on a partition $\left\{ I_j \right\}$ like for uniform expanding maps on $[0,1]$, but such that for all $j = 1, \dots , m_k$, we have $\inf\limits_{x \in I_j} |(f_k \vert_{I_j})' (x) | \ge 1$ and there exists a countable set of isolated fixed points, called \textit{indifferent fixed points}, over which the absolute value of the derivative of $f_k$ is equal to $1$.\\
A well-known example of this class of maps is given by the \textit{Pomeau-Manneville} (or \textit{Liverani-Sassuol-Vaienti}) family:
\begin{align*}
\begin{array}{l c c c}
T_\alpha : & [0,1] & \rightarrow & [0,1]\\[8pt]
\ & x & \mapsto & \lbrace x + x^{1+\alpha} \rbrace,
\end{array}
\end{align*}

where $\alpha \in \mathbb{R}^+$ and $\lbrace \cdot \rbrace$ is the fractional part function.\\
In particular, we have an indifferent fixed point for $x = 0$.\\
\\
With the following result (proved at the end of this section) we see why condition $C_d$ cannot hold for intermittent maps.

\begin{proposition}
\label{prop_noCd}
\ \\
For a fixed $k \in K$ and for every $r \in \mathbb{R}^+$, there exists a suitable $ x \in [0,1]$ such that
\begin{align*}
S_k(x) = \sum\limits_{i=1}^{+\infty} | (f_k^i)'(x) |^{-1} > r.
\end{align*}
\end{proposition}

In particular, it is impossible to find a bound for $S_k$ which is uniform in $x$.
However, an ergodic theory argument allows us to infer a lower bound on the greatest eigenvalue of the covariance matrix. It is less sharp than the result in Theorem \ref{thrm:cd}, but still highlights a polynomial nature behind the decay of the confidence region.\\

\begin{proposition}
\ \\
Consider a family $\lbrace f_k : [0,1] \rightarrow [0,1] \rbrace_{k \in K}$ of intermittent maps such that the measure $\mu_k$ is \\
\vspace{-0.4cm}

$f_k-$ergodic  and $\int\limits_{[0,1]} \max \lbrace \log |f_k'|(x),0 \rbrace d\mu_k < +\infty$.\\
Then, for every suitable pair $(x,k) \in [0,1] \times K$ and for every $N \in \mathbb{N}^*$, we have that the greatest eigenvalue $\widetilde{\lambda}_N^{(2)} (x,k)$ of the covariance matrix $\widetilde{\Gamma}_N (x,k)$ satisfies
\begin{align*}
\widetilde{\lambda}_N^{(2)} (x,k) \ge \dfrac{|L(x,k)|^2 + 1}{M^2 \left( \left[ S_k (f_k^N (x) ) \right]^2 + S_k (f_k^N (x)) N  + \dfrac{N}{6} + \dfrac{N^2}{3} \right) (N+1)} \, ,
\end{align*}
where $L(x,k) \in \mathbb{R}$ is the limit defined in Lemma \ref{lemma:limite}.
\end{proposition}

\begin{proof}
First of all, we show that $S_k (x) < +\infty.$\\
The hypotheses assumed on the family of intermittent maps allow us to apply Oseledets' Ergodic Theorem \cite{oseledets}, which grants the existence of a Lyapunov exponent $\gamma_k$ which is constant $\mu_k-$a.e. and verifies, by definition, the equality $\gamma_k = \lim\limits_{n \to +\infty}\dfrac{1}{n}\log |(f_k^n)'(x)|$, for a.e. $x \in [0,1]$.\\
\\
Using Birkhoff's Ergodic Theorem (and the ergodicity of $\mu_k$),
\begin{align*}
\begin{array}{l l}
\gamma_k & = \lim\limits_{n \to +\infty} \dfrac{1}{n} \log | (f_k^n)'(x) | = \lim\limits_{n \to +\infty} \dfrac{1}{n} \log \prod\limits_{i=0}^{n-1} |f_k'(f_k^i(x))|= \lim\limits_{n \to +\infty} \dfrac{1}{n} \sum\limits_{i=0}^{n-1} \log |f_k'(f_k^i(x))|=\\
 &= \int_{[0,1]} \log |f_k'| \, d\mu_k,
\end{array}
\end{align*}
and $|f_k'|$ equals $1$ only on a countable set of points (hence, a $\mu_k-$null set), thus $\log |f_k'|$ is positive $\mu_k-$a.e. on $[0,1]$, and so is its integral.\\
Hence, choosing, for example, the value $\dfrac{\gamma_k}{2}$, the definition of $\gamma_k$ provides a natural number $n_0=n_0(x,k) \in \mathbb{N}$ such that, for every $n > n_0$, we have
\begin{align*}
\bigg| \dfrac{1}{n} \log | (f_k^n)'(x) | - \gamma_k \bigg| \le \dfrac{\gamma_k}{2},
\end{align*}
which implies
\begin{align*}
 | (f_k^n)'(x) | \ge e^{\frac{\gamma_k}{2}n}.
\end{align*}
Therefore, if we define
\begin{align*}
c(x,k):= \sum\limits_{i=1}^{n_0} |(f_k^i)'(x)|^{-1},
\end{align*}
we find that
\begin{align*}
\begin{array}{l l}
S_k(x) & = c(x,k) + \sum\limits_{i=n_0 + 1}^{+\infty} |(f_k^i)'(x)|^{-1} \le c(x,k) + \sum\limits_{i=n_0 + 1}^{+\infty} e^{ - \frac{\gamma_k}{2}i} \le\\
\ \\
\ & \le c(x,k) + \dfrac{1}{1-e^{ - \frac{\gamma_k}{2}}} < +\infty.
\end{array}
\end{align*}
With this fact, we can proceed as in Lemma \ref{lemma:limite} and get
\begin{center}
$\sum\limits_{i=1}^{+\infty} \left\lvert \left\lvert [F_k^i (x)]^{-1} \cdot \dfrac{\partial f_k}{\partial k}(f_k^{i-1}(x)) \right\rvert \right\rvert_2 \le M S_k(x) < +\infty$,
\end{center}
which implies the existence of $L(x,k)$.\\
Now, since the derivative of intermittent is never strictly smaller that $1$, we note that
\begin{align*}
S_k(x) &= \sum\limits_{i=1}^{+\infty} |(f_k^i)'(x)|^{-1} = |f_k'(x)|^{-1} \sum\limits_{i=1}^{+\infty} | (f_k^{i-1})'(f_k(x))|^{-1} \le 1 + S_k (f_k(x)).
\end{align*}
Inductively, we find that, for every natural number $n \le N$, we may write
\begin{align*}
S_k(f_k^n(x)) \le N-n + S_k(f_k^N (x)).
\end{align*}
Following the same steps as in the proof of Theorem \ref{thrm:cd}, we find a bound on the smallest eigenvalue $\widetilde{\delta}_N^{\, (1)}(x,k)$ of the normal matrix $\widetilde{C}_N(x,k)$:
\begin{align*}
\widetilde{\delta}_N^{\, (1)}(x,k) \le \dfrac{M^2}{|L(x,k)|^2 + 1} \sum\limits_{n=0}^N \left[ S_k (f_k^n (x) ) \right]^2 \le \dfrac{M^2}{|L(x,k)|^2 + 1} \sum\limits_{n=0}^N \left[ N - n + S_k (f_k^N (x) ) \right]^2.
\end{align*}
Being
\begin{align*}
\sum\limits_{n=0}^N \left[ N - n + S_k (f_k^N (x) ) \right]^2 = \sum\limits_{n=0}^N \left[ n + S_k (f_k^N (x) ) \right]^2,
\end{align*}
and employing the well known formulas for the computation of the sum of the first $N$ natural numbers and the first $N$ squares, we get that
\begin{align*}
\widetilde{\delta}_N^{\, (1)}(x,k) \le \dfrac{M^2}{|L(x,k)|^2 + 1}\left( \left[ S_k (f_k^N (x) ) \right]^2 + S_k (f_k^N (x)) N  + \dfrac{N}{6} + \dfrac{N^2}{3} \right) (N+1),
\end{align*}
which directly leads to the proof, since $\widetilde{\lambda}_N^{(2)}(x,k) = \left[ \widetilde{\delta}_N^{\, (1)}(x,k) \right]^{-1}$.
\end{proof}

We conclude proving Proposition \ref{prop_noCd}.

\begin{proof}
Let us take an indifferent fixed point $\overline{x} \in [0,1]$ and a real number $\epsilon \in \mathbb{R}^+$ such that
\begin{align*}
\epsilon^{-1} > r.
\end{align*}
By definition of intermittent maps, we may find an open interval $I \subset [0,1]$ of the form
\begin{align*}
\begin{array}{l c l}
I=(\overline{x}, x_0 ) \ \ \ &  \text{or} \ \ \ & I=(x_0,\overline{x}),
\end{array}
\end{align*} 
such that $f_k\vert_I$ is of class $C^1$ and, for every $x \in I$,
\begin{align*}
1 < | f_k'(x) | \le 1+ \epsilon.
\end{align*}
Moreover, since $| f_k'\vert_I | > 1 $ and $f_k(\overline{x}) = \overline{x}$, we have that $f_k \vert_I : I \rightarrow f_k(I) \ \text{is invertible}$ and $I \subset f_k (I).$\\
\\
Therefore, if $x \in I$, then $| f_k'(f_k^{-1}(x)) | \le 1+\epsilon$.\\
\\
By the convergence of the geometric series of ratio less than $1$, we have
\begin{align*}
\sum\limits_{i=1}^{+\infty} (1+\epsilon)^{-i} = \dfrac{1}{1-(1+\epsilon)^{-1}} - 1 = \dfrac{1+\epsilon }{\epsilon} - 1 = \epsilon^{-1} > r.
\end{align*}
Hence, we can find $n_r \in \mathbb{N}^*$ such that
\begin{align*}
\sum\limits_{i=1}^{n_r} (1+\epsilon)^{-i} > r.
\end{align*}
We recall that, for every $i \in \mathbb{N}^*$ and for every suitable $x \in [0,1]$, \begin{center}
$|(f_k^i)'(x)| = \prod\limits_{j=0}^{i-1} |f_k' (f_k^j (x)) |$.
\end{center}
Thus, taking the $(n_r - 1)-$th counter-image $f_k^{-(n_r - 1)} (y)$ of an element $y \in I$, we can say that:
\begin{align*}
\begin{array}{l l}
S_k(f_k^{-(n_r - 1)} (y)) &= \sum\limits_{i=1}^{n_r} | (f_k^i)'(f_k^{-(n_r - 1)} (y)) |^{-1} + \sum\limits_{i=n_r +  1}^{+\infty} | (f_k^i)'(f_k^{-(n_r - 1)} (y)) |^{-1} \ge \\
\\
\ & \ge \sum\limits_{i=1}^{n_r} (1+\epsilon )^{-i} > r,
\end{array}
\end{align*}
and the proposition is proved.
\end{proof}

\section{Conclusions and Future Works}\label{section:conclusions}
%\ \\

Conditions $C_d$ and $H_d$ are among the first attempts to generalize the results in \cite{maro, maro.bonanno}, and to understand the numerical results in \cite{serra.spoto.milani, spoto.milani} through a formal mathematical framework. As such, a lot can still be done in this generalization process: though we saw that condition $C_d$ is satisfied by uniform piecewise expanding maps of the unit interval, a widely studied class of transformations, condition $H_d$ requires a certain rigidity on the derivatives of the maps, a detail which might prove unhandy. Still, the broadly known class of affine hyperbolic toral diffeomorphism represents a notable example of how condition $H_d$ can be a valid requirement.\\
Moreover, it is likely that a lot more can be found about the features of intermittent maps in this context. Our investigation is consistent with the expected behaviour, but more meaningful results in this direction wait for more detailed studies.

\end{document}